\documentclass[prb,showpacs,amsmath,amssymb,twocolumn,floatfix]{revtex4}

\usepackage{graphicx}
\usepackage{bm}

\begin{document}

\title{Collective Oscillations in Classical Nonlinear Response of a Chaotic System}

\author{Sergey V.  \surname{Malinin}$^a$}
\author {Vladimir Y. \surname{Chernyak}$^a$}
\email{chernyak@chem.wayne.edu}
\affiliation{$^a$Department of Chemistry, Wayne State University,
5101 Cass Ave,Detroit, MI 48202\\}
\date{\today}

\begin{abstract}
We consider classical response in a strongly chaotic (mixing) system. As opposed to the case of
stable dynamics, the nonlinear classical response in a chaotic system vanishes at large times. The
physical behavior of the nonlinear response is attributed to the exponential time dependence of the
stability matrix. The response also reveals certain features of collective resonances which do not
correspond to any periodic classical trajectories. We calculate analytically linear and
second-order response in a simple chaotic system and argue on the relevance of the model for
interpretation of spectroscopic data.

\end{abstract}


\maketitle

Time-domain femtosecond spectroscopy constitutes a powerful tool that probes electronic and
vibrational coherent dynamics of complex molecular systems in condensed phase
\cite{TokmakoffetalPRL,ZanniHochstrPNAS,Fayer01,Jonas03,StolowJonas}. Spectroscopic signals are
directly related to optical response functions that carry detailed information on the underlying
dynamical phenomena. At room temperatures the complexity often originates from slow strongly
anharmonic vibrational modes that can be treated within the framework of classical mechanics.

A number of studies have been devoted to the classical response in stable (integrable) dynamical
systems \cite{LeegwaterMukamel95,NEL04,KryvohuzCaoPRL05}. The nonlinear response functions have
been shown diverge linearly with time, while the divergence can be eliminated by invoking a fully
quantum description
\cite{NEL04,KryvohuzCaoPRL05}. Although integrable dynamics represents only an approximation for
realistic physical situations, weak deviations from the integrability do not eliminate the
unphysical behavior of the response functions. This follows from the fact that the divergence is
related to quasiperiodic motions on invariant tori \cite{KryvohuzCaoPRL05}. According to the
Kolmogorov-Arnold-Moser (KAM) theory, most the invariant tori are not destroyed by small
perturbations that break down integrability \cite{Arnold}. Yet, stable dynamics would be typical
for a close to equilibrium situation. At larger energies a generic situation would correspond to
dynamical behavior with chaotic features \cite{Gutzwiller}. Moreover, unstable (hyperbolic)
dynamics may be more common due to the stability of chaos with respect to perturbations. It has
been argued based on results of numerical analysis \cite{DellagoMukamel03} that chaotic dynamics
appears to observe the convergence of the classical response functions. In spite of apparent
importance, and to the best of our knowledge the problem of the nonlinear response in strongly
chaotic systems has never been addressed using analytical methods. We note that analytical
calculations of the response are rarely feasible in nonintegrable systems, whereas numerical
simulations of chaotic dynamics are complicated by the exponential divergence of stability matrices
\cite{DellagoMukamel03}.

In this Letter we show that (i) the classical response of the chaotic system exhibits decay and
oscillations as a function of times between probing pulses, and
(ii) the Fourier transform of $2D$ second order response function reveals broad and asymmetric
peaks as signatures of chaos. This is the main result or our study.

A strongly chaotic (mixing) system is characterized by a special spectrum of the Liouville operator
\cite{Ruelle86,RobertsMuz}. The spectrum consists of complex Ruelle-Pollicott (RP)  resonances that
determine the asymptotic oscillations and decay of the correlations. This yields the linear
response function directly related to the two-point correlation functions, in agreement with the
fluctuation dissipation theorem (FDT). Nonlinear response, however, turns out to be more involved
still with the noticeable effect of the resonances.

A system driven by a time-dependent external field ${\cal E}(t)$ is described by the Hamiltonian
$H_T=H-f{\cal E}(t)$ with the the function $f(\bm\eta)$ in phase space representing the dipole
(see, e.g. Ref. \onlinecite{DellagoMukamel03}).
The response functions $S^{(n)}$ that depend on $n$ time intervals describe the expansion of the
measured signal $\langle f(\bm\eta(t))\rangle=\int d\bm\eta\,\rho f$ in a functional series in
${\cal E}$. The second-order response function reads
\begin{align}
\label{second-order}
S^{(2)}(t_1,t_2)=\partial_{t_1}\int d{\bm \eta} f e^{-\hat L t_2} \left\{f,e^{-\hat L t_1}
f\partial_{E}\rho_0\right\}.
\end{align}
where $\rho_{0}$ is the equilibrium distribution and $\hat{L}=\left\{H,\cdot\right\}$ is  the
Liouville operator of the unperturbed system.

\begin{figure}[ht]
\centerline{
\includegraphics[width=2.8in, height=2in]{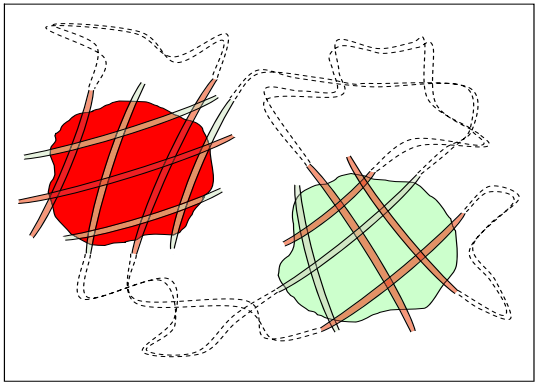}
}
\caption{
Schematic 
cross-section of the phase space along the surface given by the stable and
unstable directions.
Initial distribution of $f$ is presented by
two regions $+$ and $-$ (dark red and light green).
As time elapses the distribution elongates along unstable direction
and contracts along stable one.
\label{fettuccine}
}
\end{figure}

A schematic picture of second-order response formation in a chaotic system that employs the
Liouville phase space evolution is shown in Fig.
\ref{fettuccine}.
Propagating the observable $f$ in Eq. (\ref{second-order}) backward in time we interpret the
second-order response as an overlap of the distributions $f_{-}({\bm\eta})=\exp(\hat{L}t_{2})f$ and
$\xi^{j}\partial_{j}f_{+}({\bm\eta})$ with $f_{+}=\exp(-\hat{L}t_{1})f\partial_{E}\rho_{0}$ and the
vector field $\xi^{j}\partial_{j}=\{f,\cdot\}$.
Since $\int d{\bm x} f=0$, we can for simplicity represent the function $f$ by two separate regions
in the phase space where it adopts positive and negative values. In the case of hyperbolic dynamics
evolution during time $t$ changes the regions' shapes. For $|t|\gg 1$, the shape becomes similar to
ribbon-like fettuccine: elongated along the unstable direction by a factor $\sim e^{\lambda t}$,
narrowed along the stable one $\sim e^{-\lambda t}$ (with $\lambda$ being the Lyapunov exponent)
and unchanged along the flow. Since $f_{-}$ results from reverse dynamics, the distributions
$f_{+}$ and $f_{-}$ are elongated along the unstable and stable directions, respectively.
Therefore, the overlap is represented by a large set of $N\sim e^{\lambda(t_{1}+t_{2})}$ small
disconnected regions with the volume $v\sim e^{-\lambda(t_{1}+t_{2})}$. Since $f$ is represented by
a positive and negative regions, the distribution $f_{-}$ consists of two positively and negatively
"charged" fettuccine and the cancellations result in a signal determined by a typical fluctuation
proportional to $\sqrt{N}$, and the overlap integral attains a factor $\sqrt{N}v\sim
e^{-\lambda(t_{1}+t_{2})/2}$ that turns out to be exponentially small. Yet, this is not the end of
the story since the derivative $\xi^{j}\partial_{j}$ of a sharp feature along the stable direction
can create exponentially large $\sim e^{\lambda t_{1}}$ factors. This is the Liouville space
signature of the exponentially growing components of the stability matrix, which affects the
response starting with second order due to FDT\cite{MKC96,DellagoMukamel03}. However, the divergent
terms cancel out: decomposing ${\bm\xi}=\bm{\xi}_{0}+\bm{\xi}_{+}+\bm{\xi}_{-}$ into the
direction along the flow, and unstable and stable components, we note that only the last term is
potentially dangerous. Calculating the dangerous component of the overlap integral by parts we
arrive at two contributions: the overlap of $\xi_{-}^{j}\partial_{j}f_{-}$ with $f_{+}$, and the
overlap of $f_{-}$ with $f_{+}$ weighted with ${\rm div}{\bm\xi}_{-}$. The first contribution
contains an additional $e^{-\lambda t_{2}}$ factor, the second one provides an additional factor
independent of time. This results in a physical $\sim\exp^{-\lambda(t_{1}+t_{2})/2}$ large time
asymptotics of the nonlinear response function.

Trajectories of a particle in arbitrary potential $U({\bm r})$ are known to be the same as for a
free motion in a curved space with the metric $g_{ik}=(1-U({\bm r})/E)\delta_{ik}$
\cite{Arnold, Gutzwiller}.
Although the potential generates nonuniform motion along the trajectories, one can expect
chaotic behavior due to the exponentially growing separation between the
trajectories. In addition, when the motion is finite, the accessible part of the configurational
space at a given energy can be multiply connected. In the simplest case of two coordinates the
motion occurs inside a disk-like region punctured by $g$ forbidden islands. Some fraction of
trajectories approaches the boundaries so close that this can be qualified as reflection. Utilizing
the original argument of Sinai \cite{Sinai} reflection can be
interpreted as
continuing motion on the antipode replica of the accessible region glued to its original
counterpart via the boundaries (see Appendix \ref{app:reflection} and Ref. \onlinecite{Arnold}).
The resulting compact surface has topology of a sphere with $g$ handles (Riemann surface of genus
$g$). In the $g>1$ case the average curvature is negative, which results in unstable (hyperbolic)
dynamics.

To rationalize the described above qualitative picture of response in a chaotic system we calculate
the linear and second-order classical response functions for free motion in a Riemann surface
$M^{2}$ of constant negative (Gaussian) curvature. This model that allows an exact solution has
been serving as a prototype of classical chaos, as well as an example of semiclassical quantization
\cite{Arnold, Gutzwiller, BalazsVoros}.

The classical free-particle Hamiltonian
\begin{eqnarray}
\label{Hamiltonian} H({\bm x},\zeta)=(1/2m)g^{ik}p_i p_k=\zeta^{2}/2m\,,
\end{eqnarray}
depends on the absolute momentum value $\zeta$ only, and ${\bm x}\in M^{3}$ includes the two
coordinates and momentum direction angle $\theta$. Therefore, the smooth compact $3D$ manifold
$M^{3}$, which represents the subspace of phase space with fixed energy, is preserved in classical
dynamics. Hereafter, we will use dimensionless units so that $m=1$ the curvature $K=-1$, and
$\zeta=1$ (when energy is fixed). Our model allows for an exact solution due to strong Dynamical
Symmetry (DS). To describe DS
we adopt an agreement used in differential geometry by identifying a first-order differential
operator of differentiating along the vector field with the vector field itself. We denote by
$\sigma_{1}$ the vector field that describes the phase space dynamics, and set
$\sigma_{z}=\partial/\partial\theta$, $\sigma_{2}=[\sigma_{1},\sigma_{z}]$. A simple local
calculation yields $[\sigma_2,\sigma_z]=-\sigma_1$ and $[\sigma_{1},\sigma_{2}]=-K\sigma_{z}$ which
implies that in the constant negative curvature case  the vector fields $\sigma_{z}$, $\sigma_{1}$,
and $\sigma_{2}=[\sigma_{1},\sigma_{z}]$ form the Lie algebra $so(2,1)$.
The group $SO(2,1)$ action in the reduced phase space $M^{3}$ is obtained by integrating of the
$so(2,1)$ algebra action.
DS with respect to the action of the group $G\cong SO(2,1)$ does not mean symmetry in a usual
sense, i.e. that the system dynamics commutes with the group action, but rather reflects the fact
that the Poisson bracket is given by a bi-vector field
\begin{eqnarray}
\label{Poisson-bracket} \omega=\partial_{\zeta}\otimes\sigma_1 - \sigma_1\otimes \partial_{\zeta}
+\zeta^{-1} \left( \sigma_2\otimes\sigma_z - \sigma_z\otimes\sigma_2 \right)\,.
\end{eqnarray}
In particular, the vector field that determines classical dynamics is given by an element
$\sigma_{1}$ of the corresponding Lie algebra $so(2,1)$, whereas the stable and unstable directions
are determined by
$\sigma_z\mp\sigma_2$.

DS implies that the space of phase-space distributions constitutes a representation of $SO(2,1)$
and, being decomposed into a sum of irreducible representations, provides a set of uncoupled
evolutions. Unitary irreducible representation of $SO(2,1)$ are well-known and can be conveniently
implemented in terms of functions $\Psi(u)$ in a circle \cite{Lang}. Principal series
representations relevant for our calculations are labeled by imaginary parameter $s$ with the
Liouville operator $\sigma_{1}=\sin u\partial_{u}+(1/2-s)\cos u$. The aforementioned decomposition
identifies the angular harmonic $\Psi_k(u)=e^{iku}$ in a circle with a phase-space distribution
$\psi_k(\bm x;s)$ with given angular momentum $\sigma_{z}\psi_{k}({\bm x};s)=ik\psi_{k}({\bm
x};s)$. Commutation law in $so(2,1)$ entails $\sigma_{\pm}\psi_{k}({\bm x};s)=(\pm
k+1/2-s)\psi_{k\pm 1}({\bm x};s)$ with two anti-Hermitian conjugated ladder operators
$\sigma_{\pm}=\sigma_{1}\pm i\sigma_{2}$, $\sigma_\pm^\dag=\sigma_\mp$. Any relevant distribution
can be decomposed in the modes $\psi_{k}({\bm x};s)$, where $s$ adopts discrete values (the
spectrum). The spectrum $\{s_{\alpha}\}$ can be related to the eigenvalues of the Laplacian
operator in $M^{2}$ by identifying
$\nabla^{2}=-\frac{1}{2}(\sigma_{+}\sigma_{-}+\sigma_{-}\sigma_{+})$ and $-\nabla^{2}\psi_{0}({\bm
x};s)=(1/4-s^{2})\psi_{0}({\bm x};s)$.

The dipole $f$ is a function in $M^{2}$, i.e. a function in phase space that does not depend on
both $\zeta$ and momentum direction $\theta$, hence $\sigma_{z}f=0$, and $f$ can be expanded as a
sum over the principal series representations $f=\sum_{s}B_{s}\psi_0(\bm x;s)$.

The integrands in the response
function (\ref{second-order}) include
the result of action of the evolution operator $e^{-\hat L t}$ on the components $\psi_0(\bm x;s)$
of the dipole momentum and their derivatives resulting from the Poisson brackets.
To simplify the calculations, we transform the expressions of response functions in a way
that the evolution of derivatives of $\psi_0(\bm x;s)$ does not appear.

We use the representation in the circle to find
the expansion of
$e^{-\hat L t}\psi_0(\bm x;s)$ over basis vectors
$\psi_k({\bm x};s)$,
\begin{align}
\label{evolution-expansion}
e^{-\hat L t}\psi_0(\bm x;s)=\sum\limits_{k=-\infty}^{+\infty}A_k(t;s)\psi_k(\bm x;s)\,.
\end{align}
Since $e^{-\hat L t}\Psi_0(u)$ represents a function obtained as the result of the
action of the evolution operator on $\Psi_0(u)\equiv 1$, it can be found by solving
the equation $\partial_t g(t,u)+\hat L g(t,u)=0$ supplemented with initial condition $g(0,u)=1$.
In the principal series representation
with $\mathrm{Re}\, s=0$, $\mathrm{Im}\, s>0$, and
the equation takes the following form:
\begin{eqnarray}
\label{equation-on-circle}
\partial_t g(t,u)+\left(\sin u\partial_{u}+(1/2-s)\cos u\right)g(t,u)=0\,.
\end{eqnarray}
The solution can be easily found, and thus after the integration $\int du\, e^{-iku} g(t,u)$
we obtain $A_k(t;s)$:
\begin{align}
\nonumber
&
A_k(t;s)=
\frac{2(-1)^k\left(1-e^{-2t}\right)^k \Gamma(k+1/2-s)}
{\sqrt{\pi}\Gamma(1/2-s)}e^{-t/2}
\times
\\
&
\nonumber
{\rm Re}\left[
\frac{\Gamma(s)e^{st}}{\Gamma(k+\frac{1}{2}+s)}
\,_2F_1\left(k+\frac{1}{2}-s,k+\frac{1}{2},1-s,e^{-2t}\right)
\right],
\end{align}
where $_2F_1$ is the Gauss hypergeometric function.
For details see Appendix \ref{app:calculation} and Ref. \onlinecite{tobe}.

The linear response function can be obviously expressed via $A_0(t;s)$. The calculation involves
only the first part of the Poisson bracket, and the result is conveniently represented similarly to
the fluctuation-dissipation relation $S^{(1)}(t)\propto \partial_{t}A_0(t;s)$ \cite{MKC96}.

For large $t$ the linear response function shows damped oscillations $e^{(\pm s-1/2)t}$. The
expansion in powers of $e^{-2t}$ corresponds to RP resonances \cite{RobertsMuz}. Only even RP
resonances contribute to the response function. The expansion is a converging series in $e^{-2t}$
if $e^{-2t}<1$, i.e. $t>0$. Since the dipole $f(\bm x)$ can decomposed in irreducible
representations characterized by different values of $s$,
the linear response function constitutes a linear combination of contributions
with coefficients $B_s^2$.

The second-order response function can be calculated using Eqs. (\ref{second-order}) and
(\ref{Poisson-bracket}) \cite{tobe}. The calculation can be substantially simplified by propagating
the observable $f$ in Eq. (\ref{second-order}) backwards in time and making use of  $\left(
e^{-\hat L t_2} \right)^\dag=e^{\hat L t_2}$.
Then we apply Eq. (\ref{evolution-expansion})
to decompose $e^{\hat L t_2}\psi_0(\bm x;s)$ in $\psi_k(\bm x;s)$
and $e^{-\hat L t_1}\psi_0(\bm x;s)$ in $\psi_l(\bm x;s)$.
The integration over the reduced phase space includes an integral over the momentum direction $\theta$
that results in vanishing of all terms with $k\ne l\pm 1$. The second order response function consists
of several contributions, $S^{(2)}=\sum_j \sum\limits_{p,q,r}B_p B_q B_r\sum_{k=0}^\infty (-1)^k S^{(2)}_{j,k}$
of similar form, e.g.:
\begin{align}
\nonumber
S^{(2)}_{1,k}=
(-1)^k
k\left(k+\frac{1}{2}-p\right)a^{pqr}_k A_{k+1}^*(t_2;p)\frac{\partial A_{k}(t_1;r)}{\partial t_1 },
\end{align}
with the matrix elements
\begin{eqnarray}
\label{abqrs}
a^{s_1s_2s_3}_k=\int d\bm x\,\psi_k^*(\bm x;s_1)\psi_0(\bm x;s_2)\psi_k(\bm x;s_3)\,.
\end{eqnarray}
We further establish recurrent relations for coefficients (\ref{abqrs}) that allow expressing all
of them via few ``initial conditions'', e.g. $a^{s_1s_2s_3}_0$. For instance, in the simplest case
$s_1=s_2=s_3=s$, the recurrent relations read
$a_{k+1}=\frac{8k^2+1-s^2}{(2k+1)^2-s^2}a_k-\frac{(2k-1)^2-s^2}{(2k+1)^2-s^2}a_{k-1}$, and all
coefficients $a_k$ are expressed via $a_0$ due to the symmetry $a_k=a_{-k}$ \cite{tobe}. In this
case we find the asymptotic behavior $a_k\propto k^{-1/2\pm s}$ for $k\to\infty$. The summation
over $k$ is performed numerically. The set of coefficients $a_{0}$, as well as the spectrum
$\{s_{\alpha}\}$ are attributes of a particular Riemann surface. Constant curvature Riemann
surfaces of genus $g$ are classified by the so-called moduli spaces whose dimensions grow linearly
in $g$. Therefore, any finite set $s_{\alpha}$ of spectral elements, as well as the relevant set of
coefficients $a_{0}$ can be implemented for some particular Riemann surface, and hereafter we will
treat them as independent parameters.

The convergence of the series over $k$ is ensured by the dependence $A_k(t;s)\propto \exp(-2ke^{-t})$
for large $k$.
The series is almost sign-alternating, and the dependence of the summand magnitude on $k$ becomes
smoother with increasing $k$, independently of $t$. This allows for an effective procedure that
evaluates the series, since it does not involve numerical summation of the exponentially increasing
with $t$ number of terms. The procedure also works in any order of $e^{-t_1}$ and $e^{-t_2}$
obtained from the hypergeometric expansions of $A_k(t;s)$ \cite{tobe}.
In Ref. \onlinecite{tobe2} we extend our study of optical response by adding Langevin noise to classical
deterministic dynamics. We identify the RP resonances as the eigenvalues of the related
Fokker-Planck operator $\hat{{\cal L}}=-\kappa\nabla^{2}+\hat{L}$ in the weak-noise limit $\kappa\to
0$. We demonstrate that a standard spectral decomposition of the response functions in the
eigenmodes of the Fokker-Planck operator $\hat{{\cal L}}$ has a well-defined limit at $\kappa\to 0$
that reproduces our expression for the classical response functions.
\begin{figure}[ht]
\includegraphics[width=2.7in]{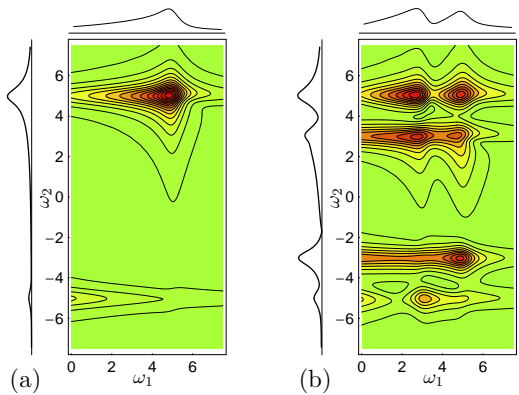}
\centerline{
}
\caption{
Absolute value of $2D$ Fourier transform of the second order response function:
(a) single resonance $s=5i$, (b) linear combination of terms with two resonances $s_1=3i$ and $s_2=5i$.
Linear plots show cross-sections of the spectra
at $\omega_1=\omega_2=5$.
\label{2dFourier}
}
\end{figure}

The absolute value of the $2D$ Fourier transform of the second-order response function presented
in Fig. \ref{2dFourier}
shows diagonal and cross peaks, as well as a stretched along the $\omega_{1}$ direction feature
(real and imaginary parts are presented in Appendix \ref{app:pictures}).
The latter originates from time-domain damped oscillations with variable period and can be interpreted
as a signature of chaos (instability) in the underlying dynamics. Also note that in our chaotic
case the peak frequencies may not be attributed to any particular periodic motions, although they
can be expressed in terms of all periodic orbits via the dynamical $\zeta$-function
\cite{Ruelle86}. Therefore, they can be referred to as collective chaotic resonances.

Our results can be applied to the interpretation of spectroscopic data. Classical chaos is quite
generic for dynamics in large molecules and molecular networks coupled via hydrogen bonds. Even for
a small number of vibrational modes, e.g. in small systems of hydrogen bonds \cite{Hbonds}, the
shape of the effective potential energy can lead to chaotic dynamics. Moreover, the chaos may be
caused by just some regions with nonuniform curvature of the potential surface \cite{Pettini}.

In the present Letter we studied classical nonlinear response in a strongly chaotic (mixing)
system. Detailed analysis performed for free motion in a Riemann compact surface of constant
negative curvature demonstrated that calculation of the response based on Liouville space dynamics
does not have fundamental difficulties. We demonstrated that peaks in $2D$ spectra are not
necessarily attributed to periodic or quasiperiodic motions, but rather can have collective nature. We
argue that collective resonances should be accompanied by stretched features in $2D$ spectra that
constitute a spectroscopic signature of underlying dynamical chaos.

\appendix

\section{Effective negative curvature of configuration space}
\label{app:reflection}

In this appendix we rationalize a close qualitative connection between classical dynamics of
multidimensional motion in a potential with forbidden islands and a geodesic flow in a
compact manifold with non-trivial topology. For the sake of simplicity we consider the case of two
coordinates. The motion is restricted to a disk-like region punctured
by forbidden islands. Boundaries of the accessible region correspond to the lines where the total
energy coincides with the potential energy.
To employ the original argument of Sinai \cite{Sinai} we glue the accessible region of the
configuration space to its antipode replica along the boundary. A reflection from the boundary can
be thought of as continuing motion in the other component. The resulting surface is compact and its
topology is characterized by genus $g$
(the number of handles attached to a sphere) corresponding to the number of original islands.
\begin{figure}[hb]
\centerline{
\includegraphics[width=2.8in]{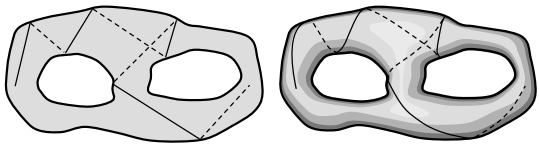}
}
\caption{ Motion in the multiply connected flat configuration space is equivalent to motion on the
curved surface. Idealization of the hard-wall potential: Trajectories are confined within the
original classically accessible $2D$ configuration space reflecting on its boundaries (left).
Reflections can be viewed as transitions to the antipode surface component glued to the original
one along the boundaries (right). Trajectories on the deformed smooth version of the resulting compact
surface are shown as solid and dashed lines, which correspond to the original and antipode
components, respectively. \label{islands} }

\end{figure}
In Fig. \ref{islands} we show both the original billiard-like motion in the configuration space and
the equivalent motion in two components assembled together that form a compact surface of genus
$g=2$. As mentioned in the main text trajectories of a particle moving in a potential resemble the
geodesic lines in  curved space with the metric $g_{ik}=(1-U({\bm r})/E)\delta_{ik}$. The regions
of the configuration space that correspond to negative curvature work as a defocusing lenses,
causing instability (divergence of close trajectories). For example, close trajectories diverge
every time they approach the boundary of the classically inaccessible island or pass a region of a
potential local maximum that belongs to the accessible region. Passing through the stable regions
with positive curvature cannot compensate for the instability, since stability reflects
oscillatory, rather than converging features in the dynamics of the trajectory deviation. In
particular, existence of unstable regions combined with ergodicity ensures exponential divergence
of trajectories over long enough times. According to the Gauss-Bonnet theorem, in the $g>1$ case
the average curvature is negative, which implies regions of instability.

\section{Liouville-space calculation of response functions}
\label{app:calculation}

Response functions can be obtained starting with the driven Liouville equation
$\partial_t\rho+\hat{\cal L}={\cal E}(t)\{f,\rho\}$. The phase space distribution $\rho(\bm\eta,
t)$ is represented as a functional series in ${\cal E}(t)$ with the initial distribution $\rho_0$
being the zero-order contribution. In our case of interest the observable coincides with the dipole
moment $f$ that represents coupling to the probe field. The linear and second-order classical
response functions are expressed via the Poisson brackets as
\begin{align}
\label{first-order0}
S^{(1)}(t) & =
\int d{\bm \eta}\, f({\bm \eta})e^{-\hat L t}\{f({\bm\eta}),\rho_0\}\,,
\\
\nonumber
S^{(2)}(t_1,t_2)
&
=\int d{\bm \eta}\,
f({\bm \eta})e^{-\hat L t_2}
\{f({\bm \eta}),e^{-\hat L t_1}\{f({\bm \eta}),\rho_0\}\}\,.
\end{align}
We start with utilizing the fact that the equilibrium distribution $\rho_0$ depends on energy
$E=\zeta^2/2$ only, and make a transformation based on the FDT. This brings us to Eq.
(\ref{second-order}).

The problem of free motion in a compact surface of constant negative curvature possesses dynamical
$SO(2,1)$ symmetry. DS means that the flow is determined by the $so(2,1)$ generator $\sigma_1$, it
also conserves the stable and unstable directions; this is expressed by the relation
$e^{\sigma_1 t}(\sigma_2\pm\sigma_z)= e^{\pm t}(\sigma_2\pm\sigma_z)e^{\sigma_1 t}$.

Any phase-space distribution can be decomposed into a direct sum of irreducible representations;
only principal series representations of $SO(2,1)$ contribute to the linear and second-order
response \cite{tobe}. DS implies that the distributions in different representations evolve
independently. Distributions in irreducible representation can be implemented as functions in a
circle with the basis of angular harmonics $\Psi_k(u)=e^{iku}$. The evolution in the circle is
determined by Eq. (\ref{equation-on-circle}) that, given the initial condition
$g(0,u)=\Psi(u)\equiv 1$, can be solved using the method of characteristics:
\begin{align}
\nonumber
g(t,u)=\left(\cosh t + \sinh t\cos u\right)^{s-\frac{1}{2}}\,.
\end{align}
Implementing the correspondence of natural scalar products in $M^3$ and in the circle, we find the
expansion coefficients in Eq. (\ref{evolution-expansion})
\begin{eqnarray}
\nonumber
A_k(t;s)=\int\limits_0^{2\pi}\frac{du}{2\pi} e^{-iku} g(t,u)\,.
\end{eqnarray}
These coefficients represent the main \textit{dynamical ingredient} in the response function
calculation. We represent them  in terms of the hypergeometric functions. Alternative
representations have been derived in the context of two-point
correlations\cite{BalazsVoros,RobertsMuz}. The linear response function is determined by the
dynamical part alone:
\begin{align}
\nonumber
S^{(1)}(t)=
\frac{\partial}{\partial t}
\int\limits_0^\infty d\zeta\, A_0(\zeta t;s)\frac{\partial \rho_0}{\partial \zeta}\,.
\end{align}

Calculation of the second-order response function is more involved. Substituting the Poisson
brackets (\ref{Poisson-bracket}) into Eq. (\ref{first-order0}) the $2D$ signal adopts a form
\begin{align}
\nonumber
&
S^{(2)}(t_1,t_2)=
\frac{\partial}{\partial t_1}
\int d{\bm x}\,d\zeta\,
\times
\\
\nonumber
&
\left\{
\zeta t_2\left(\sigma_1 e^{\sigma_1\zeta t_2}f\right)^*
(\sigma_1f)(e^{-\sigma_1\zeta t_1}f)
\right.
\\
\nonumber
&
+\left(e^{\sigma_1\zeta t_2}f\right)^*
(\sigma_1f)(e^{-\sigma_1\zeta t_1}f)
\\
\nonumber
&
\left.
+
\left( e^{\sigma_1\zeta t_2}f\right)^*
(\sigma_2f)(\sigma_z e^{-\sigma_1\zeta t_1}f)
\right\}
\frac{\partial\rho_0}{\partial\zeta}\,.
\end{align}
All terms in parentheses are further expanded in irreducible representations and angular harmonics.
The resulting contributions include two coefficients $A_k$ and a triple product given by Eq.
(\ref{abqrs}). The latter is the \textit{geometrical ingredient} that appears in nonlinear response
functions. The simplest expression for the second-order response function that incorporates
different collective resonances, originates from two terms in the decomposition
$f=B_{s_1}\psi_0(\bm x;s_1)+B_{s_2}\psi_0(\bm x;s_2)$. Then the response function reads
\begin{align}
\nonumber
&
S^{(2)}
=\int d\zeta\frac{\partial\rho_0}{\partial E}
\frac{\partial}{\partial t_1}\sum\limits_{p,q,r=s_1,s_2} B_p B_q B_r
\sum\limits_{n=0}^{\infty}
(-1)^n
\times
\\
&
\nonumber
\Biggl\{
\biggl[\biggl(n+\frac{1}{2}+r\biggr)a^{pqr}_n-\biggl(n+\frac{1}{2}+p\biggr)a^{pqr}_{n+1}\biggr]
\times
\\
&
\nonumber
\biggl[t_2\frac{\partial}{\partial t_2}-n\biggr]( A_{n}^*(t_2;p)A_{n+1}(t_1;r))
\\
&
\nonumber
-\biggl[\biggl(n+\frac{1}{2}-p\biggr)a^{pqr}_n-\biggl(n+\frac{1}{2}-r\biggr)a^{pqr}_{n+1}\biggr]
\times
\\
&
\nonumber
\biggl[t_2\frac{\partial}{\partial t_2}+n+1\biggr]( A_{n+1}^*(t_2;p)A_{n}(t_1;r))
\Biggr\}
\,.
\end{align}
It turns out that the resulting second-order response function may be represented as a double
expansion over RP resonances. The expansion is nontrivial: one encounters diverging series if the
expansion of quantities $A_n(t;s)$ is performed first. Fortunately, we have found an efficient
scheme for numerical evaluation of the series\cite{tobe}. The procedure also demonstrates that the
second-order response is indeed expanded in the RP resonances.

\section{2D spectra for spectroscopy of a chaotic system}
\label{app:pictures}

\begin{figure}[ht]
\centerline{
\includegraphics[width=2.55in]{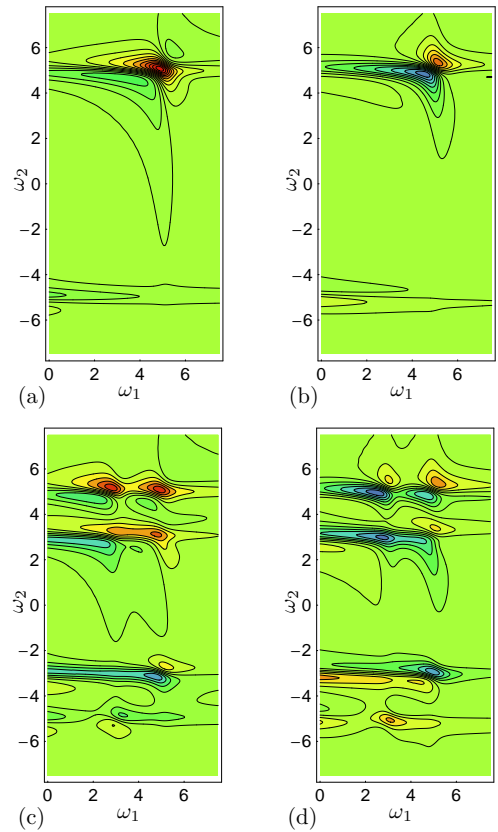}
} \caption{ Real and imaginary parts of $2D$
spectra: (a) real part, (b) imaginary part with the single resonance $s=5i$; (c) real part, (d)
imaginary part of linear combination of terms with two different resonances $s_1=3i$ and $s_2=5i$.
\label{realimaginary} }
\end{figure}

In this appendix we present some details on $2D$ spectra that characterize the second-order
response of a chaotic dynamical system considered in this Letter. Experimental data on $2D$
time-domain spectroscopy that probes the response function $S^{(2)}(t_{1},t_{2})$ is usually
presented using the so-called $2D$ spectra that constitute a numerical $2D$ Fourier transform of
the response function with respect to $t_{1}$ and $t_{2}$. The absolute value of the $2D$ spectra
for our dynamical system are given in Fig. \ref{2dFourier}, whereas in Fig. \ref{realimaginary} we
present the real and imaginary parts.

As follows from the general picture presented in the main text, for a given unperturbed dynamics
the response function depends on the expansion of the dipole function $f$ in the eigenmodes of the
Laplace operator: each eigenmode provides with an oscillation in the signal. Since the dipole is
generally a smooth function of the system coordinates, the number of eigenmodes that participate in
the expansion is typically small. To study the important features of the signals we consider the
cases of one and two oscillation in the signal. In the first case we see a diagonal peak with a
stretched feature along $\omega_{1}$ direction; such stretching never occurs due to a periodic
orbit in an integrable system even if it is coupled to a harmonic bath. In the second case we see
both diagonal and cross peaks accompanied by the stretched features.


\begin{thebibliography}{99}

\bibitem{TokmakoffetalPRL} A. Tokmakoff,  M. J. Lang, D. S. Larsen, G. R. Fleming, V. Chernyak,
and S. Mukamel, Phys. Rev. Lett. {\bf 79}, 2702 (1997).

\bibitem{ZanniHochstrPNAS} M. C. Asplund, M. T. Zanni, and R. M. Hochstrasser,
Proc. Natl. Acad. Sci. USA {\bf 97}, 8219 (2000).

\bibitem{Fayer01} M. D. Fayer, Ann. Rev. Phys. Chem. {\bf 52}, 315 (2001).

\bibitem{Jonas03} D. M. Jonas, Ann. Rev. Phys. Chem. {\bf 54}, 425 (2003).

\bibitem{StolowJonas} A. Stolow and D. M. Jonas, Sciense {\bf 305}, 1575 (2004).

\bibitem{LeegwaterMukamel95} J. A. Leegwater and S. Mukamel, J. Chem. Phys. {\bf 102}, 2365 (1995).

\bibitem{NEL04} W. G. Noid, G. S. Ezra, and R. F. Loring, J. Phys. Chem.  B {\bf 108}, 6536 (2004).

\bibitem{KryvohuzCaoPRL05} M. Kryvohuz and J. Cao, Phys. Rev. Lett. {\bf 95}, 180405 (2005);
{\bf 96}, 030403 (2006);
J. Chem. Phys.{\bf 122}, 024109 (2005).


\bibitem{Gutzwiller} M.~C.~Gutzwiller, {\it Chaos in Classical and Quantum Mechanics},
Springer Verlag, 1990.

\bibitem{Arnold} V.~I.~Arnold, {\it Mathematical Methods of Classical Mechanics},
Springer Verlag, 1989.

\bibitem{BalazsVoros} N. L. Balazs and A. Voros, Phys. Rep. {\bf 143}, 109 (1986).

\bibitem{Sinai} Y. G. Sinai, Russ. Math. Surv. {\bf 25}, 137 (1970).

\bibitem{DellagoMukamel03} C. Dellago and S. Mukamel, Phys. Rev. E {\bf 67}, 035205(R) (2003);
J. Chem. Phys. {\bf 119}, 9344 (2003).

\bibitem{Ruelle86} D. Ruelle, Phys. Rev. Lett. {\bf 56}, 405 (1986).

\bibitem{RobertsMuz} S. Roberts and B. Muzykantskii, J. Phys. A: Math. Gen. {\bf 33}, 8953 (2000).

\bibitem{MKC96} S. Mukamel, V. Khidekel, and V. Chernyak, Phys. Rev. E {\bf 53}, R1 (1996).

\bibitem{Martin} P. C. Martin, \textit{Mesurements and correlation functions}
(Gordon and Breach, New York, 1968).

\bibitem{Kirillov} A.~A.~Kirillov, {\it Elements of the Theory of Representation of Groups},
Springer Verlag, 1986.

\bibitem{Lang} S.~Lang, $SL_{2}(R)$, Addison-Wesley, 1975.

\bibitem{Williams} F.~L.~Williams, {\it Lectures on the Spectrum of} $L^{2}(\Gamma\backslash G)$.

\bibitem{tobe} S. V. Malinin and V. Y. Chernyak, to be published.

\bibitem{tobe2} S. V. Malinin and V. Y. Chernyak, to be published.

\bibitem{Hbonds} J. D. Eaves, J. J. Loparo, C. J. Fecko, S. T. Roberts, A. Tokmakoff,
and P. L. Geissler, Proc. Natl. Acad. Sci. USA {\bf 102}, 13019 (2005).

\bibitem{Pettini} L. Casetti, C. Clementi, and M. Pettini, Phys. Rev. E {\bf 54}, 5969 (1996).

\end{thebibliography}
\end{document}